\documentclass[aps, preprint, showpacs, amsmath, amssymb, prl, 12pt]{revtex4}
\usepackage{graphicx}
\usepackage{dcolumn}
\usepackage{bm}
\begin{document}
\title{Strongly Correlated Two-Electron Transport in a Quantum Waveguide Having a  Single Anderson Impurity}
\author{Jung-Tsung Shen}
\email{jushen@stanford.edu}
\author{Shanhui Fan}
\email{shanhui@stanford.edu}
\affiliation{Ginzton Laboratory, Stanford University, Stanford, CA
94305}
\date{\today}
\begin{abstract}
The strongly correlated two-electron transport in one-dimensional channel coupled with an Anderson-type impurity is solved exactly via a Bethe ansatz approach. We show that the transport properties are fundamentally different for spin singlet and triplet states, thus the impurity acts as a novel filter that operates based on the total spin angular momentum of the electron pairs, but not individual spins. The filter provides a deterministic generation of electron entanglement in spin, as well as energy and momentum space.
\end{abstract}
 \maketitle

There has been a significant interest in creating controllable entanglement between two electrons, for the purpose of creating electron-based solid-state systems for quantum information processing~\cite{Burkard:1999, Oliver:2002, Costa:2001, Lebedev:2004, Burkard:2000, Recher:2002}. Among many proposals to generate entanglement between electrons, the one of mobile electron scattering in a one-dimensional channel is of particular interest. In such scattering processes, the strong correlations between flying electrons could be induced either via direct Coulomb interactions in quantum dots~\cite{Oliver:2002}, 
or by a localized magnetic impurity~\cite{Costa:2001}. 
This type of proposal is attractive due to its deterministic generation of entanglement, without the need of additional processes such as post-selection~\cite{Lebedev:2004}.
However, in all previous works, the treatment is perturbative, and yet it is well-known that these systems have interesting properties beyond perturbation theory.  

In this Letter, we consider the strongly correlated two-electron transport in one-dimensional channel coupled with an Anderson-type empty orbital. We show that this problem can be solved exactly via a Bethe ansatz approach. Our analysis shows that the transport properties are fundamentally different for spin singlet and triplet states. One of the most remarkable results out of this exact solution is that, for the singlet state, the strong on-site Coulomb repulsion can generate spatial electron-electron pairing via the scattering process. In addition, in the singlet case, the energy spectra of the transmitted and reflected scattering states of the two entangled electrons exhibit a continuous distribution with side bands, a phenomenon analogous to the photonic resonance fluorescence~\cite{Mollow:1969}. Furthermore, the transmitted two-electron state can manifest either bunching or anti-bunching behavior, in contrast to the case considered in Ref.~[\onlinecite{Burkard:2000}] wherein the singlet states only show bunching behavior. This indicates that the spatial correlation between entangled electrons is a tunable dynamical process. On the other hand, in the triplet case, the transport properties of the two electron are the same as those of two independent electrons, with no spatial electron-electron pairing. Finally, we show that based upon the different behaviors of the singlets and the triplets, the system acts as a novel filter that operates based on the total spin angular momentum of the electron pairs, but not individual spins, and that the filter provides a deterministic generation of electron entanglement in spin, as well as energy and momentum space.

\begin{figure}[thb]
\scalebox{1}{\includegraphics[width=\columnwidth]{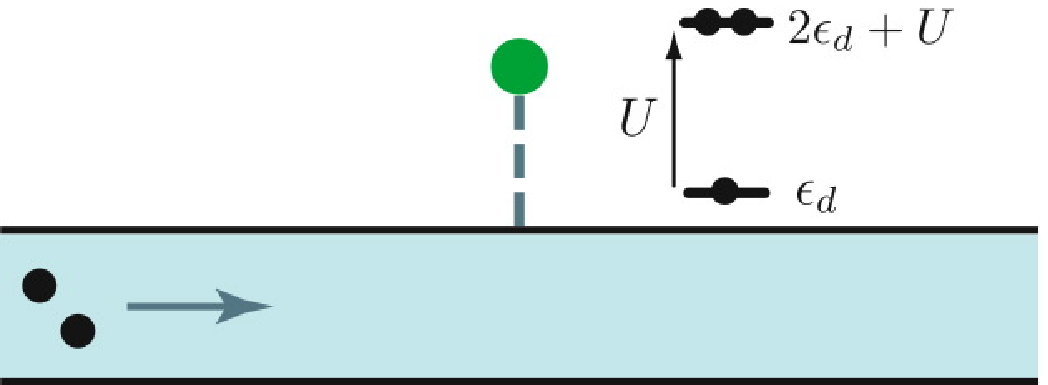}}
\caption{(color online). Schematics of the system. Two electrons (black dots) propagate in a one-dimensional channel (light blue) coupled to a quantum impurity (green dot). The coupling is indicated by dash line. Also shown is the energy diagram of the quantum impurity: when only one electron occupies the impurity, the energy is $\epsilon_d$; when two electrons with opposite spins occupy the impurity, the total energy is $2\epsilon_d + U$.}\label{Fi:Schematics}
\end{figure}

To begin with, we set up the system as shown in Fig.~\ref{Fi:Schematics}. The electrons freely propagate in the one-dimensional channel which couples to an Anderson-type empty orbital~\cite{Anderson:1961}. In real-space, the Anderson Hamiltonian takes the following form~\cite{Wiegmann:1983a, Hewson:1997}:
\begin{align}\label{E:AndersonTwoMode}
H =& \sum_{\sigma=\uparrow, \downarrow} \int\left\{-i v_g c^{\dagger}_{R\sigma}(x)\frac{\partial}{\partial x}c_{R\sigma}(x)+ i v_g c^{\dagger}_{L\sigma}(x)\frac{\partial}{\partial x}c_{L\sigma}(x) \right.\notag\\
&+ \left.\bar{V} \delta(x)\left(c^{\dagger}_{R\sigma}(x) d_{\sigma} + c_{R\sigma}(x)  d^{\dagger}_{\sigma} +  c^{\dagger}_{L\sigma}(x) d_{\sigma} + c_{L\sigma}(x) d^{\dagger}_{\sigma} \right)\right\}\notag\\
&+\sum_{\sigma=\uparrow, \downarrow}\epsilon_{d} n_{d,\sigma} + U n_{d\uparrow} n_{d\downarrow}
\end{align}
The first two terms describe the kinetic energy of the right-moving ($R$) and left-moving ($L$) electrons in the one-dimensional continuum, respectively, where $v_g$ is the group velocity. $c^{\dagger}_{R, \sigma}(x)$ ($c_{R, \sigma}(x)$) is the creation (annihilation) operator of the right-moving electron with spin $\sigma$, and similarly for $c^{\dagger}_{L, \sigma}(x)$ ($c_{L, \sigma}(x)$). The third term describes the processes of electron hopping on and off the impurity at $x=0$ (``hybridization''), where $\bar{V}$ is the coupling strength, and the operator $d^{\dagger}_{\sigma}$ ($d_{\sigma}$) creates (annihilates) an electron of spin $\sigma$ on the impurity. $n_{d,\sigma} \equiv d^{\dagger}_{\sigma} d_{\sigma}$ is the electron number operator on the impurity. $\epsilon_d$ is the energy of the empty orbital of the impurity, degenerate for both spins of the electron. The last term with $U>0$ is the energy cost to put two electrons with opposite spin on the same impurity orbital. Double occupation of electrons is energetically unfavorable, and becomes prohibited in the limit $U \rightarrow +\infty$. 
The Hamiltonian of Eq.~(\ref{E:AndersonTwoMode}) describes the situation where the electrons can propagate in both directions, and is referred to as ``two-mode'' model. 

By employing the transformation $c^{\dagger}_{e\sigma}(x)\equiv (c^{\dagger}_{R\sigma}(x)+ c^{\dagger}_{L\sigma}(-x))/\sqrt{2}$, and $c^{\dagger}_{o\sigma}(x)\equiv
(c^{\dagger}_{R\sigma}(x)- c^{\dagger}_{L\sigma}(-x))/\sqrt{2}$, the original two-mode Hamiltonian is transformed into two decoupled ``one-mode''
Hamiltonians describing an even and an odd subspaces, \emph{i.e.,} $H=H_e + H_o$:
\begin{align}\label{E:HeHo}
H_e &= \sum_{\sigma}\int dx (-i) v_g c_{e\sigma}^{\dagger}(x)\frac{\partial}{\partial x}
c_{e}(x) + \int dx V \delta(x)\left(c^{\dagger}_{e\sigma}(x)
\sigma_{-} + c_{e}(x) \sigma_{+}\right)
+\sum_{\sigma}\epsilon_{d} n_{d,\sigma} + U n_{d\uparrow} n_{d\downarrow},\notag\\
H_o &= \sum_{\sigma}\int dx (-i) v_g c_{o\sigma}^{\dagger}(x)\frac{\partial}{\partial x}
c_{o\sigma}(x),
\end{align}with $[H_e, H_o]=0$. $H_o$ is an interaction-free  Hamiltonian, while
$H_e$ describes a non-trivial interacting model with coupling strength $V\equiv\sqrt{2} \bar{V}$. $H_e$ has also been used to describe the S-wave scattering of electrons off a single magnetic impurity in three dimensions. Also, with the substitution $V \rightarrow \sqrt{|t_{L}|^2 + |t_{R}|^2}$, $H_e$  describes the one-dimensional transport problem of electrons tunneling into a small quantum dot, with $t_{L}$, $t_{R}$ being the tunneling amplitude from the left lead and right lead, respectively~\cite{Bruus:2004}. For notational simplicity, $v_g$ and $\hbar$ are set to 1 hereafter.

The transport properties is fully encoded in the S-matrix, $\mathbf{S}$. For a prepared \emph{free} incoming state (in-state) $|\Psi_\text{in}\rangle$, the free state describing the   outcome (out-state) is given by $|\Psi_\text{out}\rangle = \mathbf{S} |\Psi_\text{in}\rangle$~\cite{Taylor:1972, Shen:2007d}.  Since both $c^{\dagger}_{R\sigma}(x)$ and $c^{\dagger}_{L\sigma}(x)$ are linear combinations of $c^{\dagger}_{e\sigma}(x)$ and $c^{\dagger}_{o\sigma}(x)$, 
any \emph{free} two-electron state $|\Psi_2\rangle$ can be written as 
\begin{equation}
|\Psi_2\rangle = |\Psi\rangle_{ee} + |\Psi\rangle_{eo} + |\Psi\rangle_{oe} + |\Psi\rangle_{oo},
\end{equation} where the subscript $ee$, for example, labels the subspace spanned by $c^{\dagger}_{e\sigma}(x_1)c^{\dagger}_{e\sigma'}(x_2)|\emptyset\rangle$. (The spin labels are suppressed for simplicity.) Moreover, since the Hamiltonians, Eq.~(\ref{E:AndersonTwoMode}) and (\ref{E:HeHo}), do not mix the $e$ and $o$ subspaces, the following \emph{decomposition relation} holds~\cite{Shen:2007d}: 
\begin{equation}\label{E:Decomposition}
\mathbf{S} |\Psi_{2}\rangle = \mathbf{S}_{ee} |\Psi\rangle_{ee}+ \mathbf{S}_{eo} |\Psi\rangle_{eo} + \mathbf{S}_{oe} |\Psi\rangle_{oe} + \mathbf{S}_{oo} |\Psi\rangle_{oo},
\end{equation}where $\mathbf{S}_{ee}$ is the two-electron S-matrix in the $ee$ subspace governed by $H_e$; $\mathbf{S}_{eo} =\mathbf{S}_{oe} = \mathbf{S}_{e} \mathbf{S}_{o}$, with $\mathbf{S}_{e}$ being the one-electron S-matrix in the $e$ subspace, and $\mathbf{S}_{o} = \mathbf{1}$, the identity operator, being the one-electron S-matrix in the $o$ subspace governed by $H_o$;  $\mathbf{S}_{oo} = \mathbf{1}$. 

Thus, for a given two-electron in-state $|\Psi_2\rangle$, to compute the out-state $\mathbf{S}|\Psi_{2}\rangle$, one first decomposes the in-state $|\Psi_{2}\rangle$ into  $ee$, $oo$, $eo$, and $oe$ subspaces, followed by computing the scattering states in each subspace, and finally transforms the results back to the original $RR$, $LL$, $RL$, and $LR$ spaces, using Eq.~(\ref{E:Decomposition}). Note that the Hamiltonian in Eq.~(\ref{E:AndersonTwoMode}) conserves the total spins of the two electrons, thus the singlet and the triplet states do not mix with each other by scattering. Below we treat these cases separately. 

\emph{Singlet.} Any two-electron spin singlet eigenstates of the Hamiltonian $H_e$ is of the form~\cite{Wiegmann:1983a}:
\begin{align}
|\Phi\rangle \equiv& \left\{ \int dx_1 dx_2 \, g(x_1, x_2) c^{\dagger}_{e\uparrow}(x_1) c^{\dagger}_{e\downarrow}(x_2) + \int dx \, e(x) \left[c^{\dagger}_{e\uparrow}(x) d^{\dagger}_{\downarrow} - c^{\dagger}_{e\downarrow}(x) d^{\dagger}_{\uparrow}\right] + f d^{\dagger}_{\uparrow}d^{\dagger}_{\downarrow}\right\}|\emptyset\rangle,
\end{align} where $g(x_1, x_2)$, $f$ and $e(x)$ can be analytically solved via Bethe ansatz (see Refs.~[\onlinecite{Wiegmann:1983a, Shen:2007d}]).

In order to construct the S-matrix, one interprets $g(x_1, x_2)$ in the third quadrant ($x_1, x_2 <0$) as the in-state, and in the first quadrant ($x_1, x_2 >0$) as the out-state. This interpretation can be rigorously proved using Lippmann-Schwinger formalism~\cite{Shen:2007d}. By direct reading off the results from Ref.~[\onlinecite{Wiegmann:1983a}], one has
\begin{equation}\label{E:gInFirstThird}
g(x_1, x_2)\equiv
\begin{cases}
	\phantom{}_{ee}\langle x_1, x_2|W_{k,p}\rangle_{ee}, &\text{$x_1, x_2 < 0$;}\\
	t_k t_p \,\phantom{}_{ee}\langle x_1, x_2|W_{k,p}\rangle_{ee}, &\text{$x_1, x_2 > 0$;}
\end{cases}
\end{equation}
where $t_k \equiv(k - \epsilon_{d} - i \Gamma/2)/(k -\epsilon_{d} + i \Gamma/2)$ with $\Gamma\equiv V^2$ is the one-mode single electron transmission amplitude, and similar for $t_p$.  \begin{equation}\label{E:WkpDef}
|W_{k,p}\rangle_{ee} \equiv \frac{(k-p)(E-2\epsilon_d -U)|S_{k,p}\rangle_{ee} - i U \Gamma |A_{k,p}\rangle_{ee}}{\sqrt{(k-p)^2 (E-2\epsilon_d -U)^2 +U^2\Gamma^2}},
\end{equation}with
\begin{align}
\langle x_1, x_2|S_{k, p}\rangle_{ee} &\equiv S_{k, p}(x_1, x_2) = \frac{e^{i k x_1} e^{i p x_2} +e^{i k x_2} e^{i p x_1}}{2\pi\sqrt{2}},\notag\\
\langle x_1, x_2|A_{k,p}\rangle_{ee} &\equiv A_{k, p}(x_1, x_2) = \frac{\text{sgn}(x)\left(e^{i k x_1} e^{i p x_2} -e^{i k x_2} e^{i p x_1}\right)}{2\pi\sqrt{2}},
\end{align} and normalized as $\phantom{}_{ee}\langle W_{k', p'}|W_{k,p}\rangle_{ee} = \delta(k-k')\delta(p-p')$. As a result, we have $\mathbf{S}_{ee}|W_{k,p}\rangle_{ee}=t_k t_p |W_{k,p}\rangle_{ee}$. 

To describe the entire scattering properties, we would need to ensure that we provide the mapping for all states in the free two-electron Hilbert space. Thus, a completeness check on $\{|W_{k,p}\rangle_{ee}\}$ is crucial. Remarkably,  we found that the completeness depends upon the sign of $E-2\epsilon_d -U$. When $E-2\epsilon_d -U \geq 0$, the set $\{|W_{k, p}\rangle_{ee}: k\leq p\}$ alone forms an orthonormal complete set of basis that spans the free two-electron Hilbert space. However, when $E-2\epsilon_d -U \leq 0$, an additional \emph{two-electron bound state} $|B_E\rangle_{ee}$, defined as
\begin{equation}\label{E:BDef}
\langle x_1, x_2 |B_E\rangle_{ee} \equiv B_{E}(x_1, x_2) = \frac{\sqrt{\Gamma'}}{\sqrt{4 \pi}} e^{i E x_c -\Gamma' |x|/2},
\end{equation} is needed, such that $\{|W_{k\leq p}\rangle_{ee},  |B_E\rangle_{ee}\}$ together forms a  complete and orthonormal basis set. Here, $\phantom{}_{ee}\langle B_{E'}|B_{E}\rangle_{ee} = \delta(E-E')$, and $\phantom{}_{ee}\langle W_{k, p}|B_{E}\rangle_{ee} =0$.  $\Gamma' \equiv (U\Gamma)/(U+2\epsilon_d -E) >0$ defines the effective size of the bound state $|B_E\rangle_{ee}$. Such bound state has been noted in other context before~\cite{Kawakami:1981}. We show here that including it is crucial for describing the transport properties. 

The S-matrix for the singlet state in the $ee$ subspace, $\mathbf{S}_{ee}$, thus can be diagonalized as: 
\begin{equation}\label{E:SingletSeeMatrix}
\mathbf{S}_{ee}\equiv
\begin{cases}
	\sum_{k \leq p}  t_k t_p |W_{k,p}\rangle \langle W_{k, p}|, &\text{$E-2\epsilon_d -U \geq 0$;}\\
	\sum_{k \leq p}  t_k t_p |W_{k,p}\rangle \langle W_{k, p}| + \sum_{E} t_E |B_E\rangle\langle B_E|, &\text{$E-2\epsilon_d -U < 0$;}
\end{cases}
\end{equation} where  
\begin{equation}
t_E \equiv \frac{(E-2\epsilon_d -U)(E-2\epsilon_d-2 i \Gamma)-\Gamma^2}{(E-2\epsilon_d -U)(E-2\epsilon_d+2 i \Gamma)-\Gamma^2},
\end{equation}is the transmission amplitude for the two-electron bound state as a whole. 
The two-electron bound state, $|B_{E}\rangle$, has a spatial extent of $1/\Gamma'$. It behaves as an effective single composite particle with an energy $E=k+p$, and remains integral when passing through the impurity, acquiring a phase shift $t_E$. By tuning $\Gamma'$ via $\epsilon_d$ or $U$, the quantum impurity therefore provides a local means of manipulating composite particles of electrons. The other three remaining terms in Eq.~(\ref{E:Decomposition}) can be easily computed~\cite{Shen:2007d}, and altogether, Eq.~(\ref{E:Decomposition}) provides a complete description of the transport properties.

Using the S-matrix, Eq.~(\ref{E:SingletSeeMatrix}), we now compute the transport properties for the singlet state. Consider a singlet in-state
\begin{equation}\label{E:InSingletRR}
|\Psi_\text{in}\rangle^{\text{\scriptsize s}} \equiv |S_{k_1,p_1}\rangle_{R\uparrow R\downarrow} =\int dx_1 dx_2 \,S_{k, p}(x_1, x_2) c_{R\uparrow}^{\dagger}(x_1) c_{R\downarrow}^{\dagger}(x_2) |\emptyset\rangle
\end{equation}
incident from the left. 
The out-state, $ \mathbf{S} |\Psi_\text{in}\rangle^{\text{\scriptsize s}}$, gives the amplitudes for all possible outcomes: both electrons transmitted ($t_2^{\text{\scriptsize s}}(x_1, x_2) \equiv \langle \emptyset| c_{R\downarrow}(x_2) c_{R\uparrow}(x_1)|\Psi_\text{out}\rangle^{\text{\scriptsize s}}$), both reflected, and one electron transmitted and one reflected. All of these amplidutes have exact analytic forms. Below, we focus on $t_2^{\text{\scriptsize s}}(x_1, x_2)$, which contains the spatial correlation properties of the two transmitted electrons.

Using Eq.~(\ref{E:Decomposition}) and (\ref{E:SingletSeeMatrix}), one has~\cite{Shen:2007d}
\begin{equation}\label{E:t2SingletWavefunction}
t_2^{\text{\scriptsize s}}(x_1, x_2) =\bar{t}_k \bar{t}_p S_{k,p}(x_1, x_2) + \frac{1}{4}\sum_{\Delta' \leq 0} B S_{E, \Delta'}(x_1, x_2),
\end{equation}where 
$\bar{t}_{k,p}\equiv (t_{k,p} +1)/2$ are the two-mode single-electron transmission amplitudes, and $\Delta\equiv (k-p)/2$ is the energy difference between the two electrons. $B$ represents the resonance fluorescence similar to that in quantum optics~\cite{Mollow:1969,Shen:2007a}. When $E-2\epsilon_d - U <0$,  $B$ is given by
\begin{align}\label{E:B}
&B =
\frac{U}{U-\delta E - i\Gamma}\left(\frac{16 i \Gamma^2}{\pi}\frac{\delta E + i\Gamma}{\left[4\Delta^2 -(\delta E + i\Gamma)^2\right] \left[4{\Delta'}^2 -(\delta E + i\Gamma)^2\right]}\right)\notag\\
&-\left\{\frac{32 i U^4 \Gamma^6 (U-\delta E)^2}{\pi(U-\delta E -i\Gamma)}\frac{1}{\left[4\Delta^2 (U -\delta E)^2 +U^2 \Gamma^2\right] \left[4{\Delta'}^2 (U -\delta E)^2 +U^2 \Gamma^2\right]}\right.\notag\\
&\left.\times\frac{1}{\left[\delta E (\delta E-U)+i\Gamma(\delta E - 2U)\right]\left[\delta E(\delta E-U)+2 i\Gamma(\delta E - U)-\Gamma^2\right]}\right\},
\end{align} where $\delta E \equiv E - 2\epsilon_d$. When $E-2\epsilon_d -U \geq 0$, $\{|W_{k\leq p}\rangle_{ee}\}$ alone is a complete set of basis, one can similarly obtain a different expression for $B$ and $t_2^{\text{\scriptsize s}}$ that does not contain the contributions from $|B_E\rangle_{ee}$~\cite{Unpublished}.
The background fluorescence indicates that only the total energy, but not the individual energy of each electron, is conserved. Thus, as one electron inelastically scattering off a transient composite object formed by the quantum impurity and the other electron, the individual energy of each electron is redistributed over a continuous range, described by $B$~\cite{Shen:2007a, Shen:2007d}. The background fluorescence arises purely from on-site interaction. For the limiting case $U\rightarrow 0$, it completely disappears. 

We now explain Eq.~(\ref{E:t2SingletWavefunction}). The first term in Eq.~(\ref{E:t2SingletWavefunction}) describes independent transport of electrons, while the second term describes correlated transport. In general, the effect of correlation occurs when $x_1 \simeq x_2$. Since the poles for $\Delta$ in $B$ are all complex, the correlated term decays to zero when $|x_1 - x_2| \gg 0$. The poles in $B$ also indicate single-electron and two-electron resonances. Specifically, when $|\delta E| \simeq  2|\Delta|$, one electron has energy close to $\epsilon_d$ and is on resonance with the impurity. On the other hand, when $\delta E \simeq U \neq 0$, the \emph{electron pair} is on resonance with the impurity. The correlation typically attains its maximum degree when both conditions are simultaneously satisfied. 

\begin{figure}[thb]
\scalebox{1}{\includegraphics[width=\columnwidth]{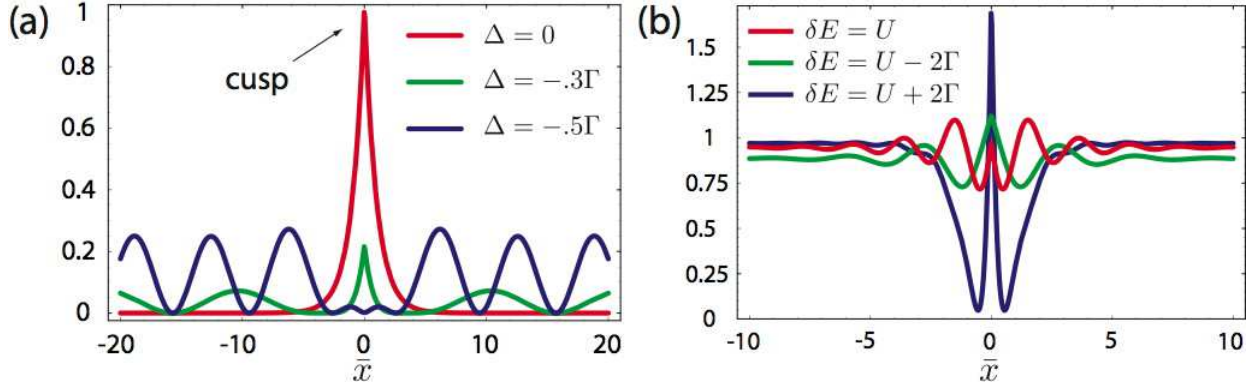}}
\caption{(color online). Singlet $|t_2^{\text{\scriptsize s}}(x_1, x_2)|^2$ (normalized by $(\sqrt{2}/2\pi)^2$). $U = 6 \Gamma$ for both cases. $\bar{x}\equiv \Gamma (x_1 - x_2)$. (a) $\delta E = 0$. The wavefunction transitions from bunching ($|t_2^{\text{\scriptsize s}}(\bar{x}=0)|^2 \simeq 1$, when $\Delta= 0$) to antibunching ($|t_2^{\text{\scriptsize s}}(\bar{x}=0)|^2 = 0$ when $\Delta=-0.5\Gamma$). (b) $\Delta = 0$.}\label{Fi:t2}
\end{figure}

We first look at the effects due to the presence of only single-particle resonance. Fig.~\ref{Fi:t2}(a) plots $|t_2^{\text{\scriptsize s}}(x_1, x_2)|^2$ for the case $\delta E=0$ and $U=6\Gamma$. The transmitted singlet state transitions from \emph{bunching} to \emph{antibunching} when $\Delta$ varies from $0$ to $-0.5\Gamma$. Thus, the bunching behavior of the singlet state is a dynamical process depending upon the tunable parameters of the system, and could be tuned via a gate voltage which changes $U$ or $\epsilon_d$. Note also that when $U \rightarrow \infty$, $B$ is greatly simplified, and the scattering properties are in fact equivalent to the strongly correlated two-photon transport of one-dimensional Dicke model~\cite{Shen:2007a, Shen:2007d}.

For the special case where both electrons satisfy the single-electron resonance condition, \emph{i.e.} $E=2\epsilon_d$ and $\Delta=0$,  Eq.~(\ref{E:t2SingletWavefunction}) yields
\begin{equation}\label{E:t2Integrated}
t_2^{\text{\scriptsize s}}(x_1, x_2)  = \frac{\sqrt{2}}{2\pi}e^{i E x_c}\left[-\left(1+\frac{U\Gamma}{(U-i\Gamma)(2iU+\Gamma)}\right)e^{-\Gamma |x|/2}\right],
\end{equation}as shown in Fig.~\ref{Fi:t2}(a).
$|t_2^{\text{\scriptsize s}}(x_1, x_2)|^2$ decays exponentially as $|x|\equiv |x_1-x_2|$ becomes large, and thus the two transmitted electrons form a \emph{bound state}.  When $|x|$ is small, $|t_2^{\text{\scriptsize s}}|^2 \propto 1-\Gamma |x|$ shows a cusp at $x=0$, so is $|r_2|^2$. These features should manifest in the measurement of the $g^{(2)}(\tau)$ function in each case. 

Fig.~\ref{Fi:t2}(b) plots $|t_2^{\text{\scriptsize s}}(x_1, x_2)|^2$ for the case when only two-electron resonance manifests, with $\delta E \simeq U=6\Gamma$ and $\Delta=0$. In this case, away from $x=0$, since neither electron is at the single-electron resonance, the wavefunction approaches that of two independent electrons for $\Delta=0$, with a near-unity transmission coefficient. In the vicinity of $x=0$, the electrons can exhibit a strong bunching effect, but no strong anti-bunching is observed.

\emph{Triplet.} The triplet cases ($S=1$, $S_z=0, 1$) can be solved straightforwardly in the same fashion. Here we will summarize the key findings: (1) $f\equiv0$, which indicates that double occupation of electrons at the impurity are always prohibited; and (2) independent of $E$, the set $\{|T_{k, p}\rangle_{ee}: k\leq p\}$, with $|T_{k, p}\rangle_{ee}$ defined as
\begin{equation}\label{E:tbaseDef}
\langle x_1, x_2|T_{k, p}\rangle_{ee} \equiv T_{k, p}(x_1, x_2)= \frac{e^{i k x_1} e^{i p x_2} -e^{i k x_2} e^{i p x_1}}{2\pi\sqrt{2}},
\end{equation} alone forms an orthonormal complete set in the free two-electron Hilbert space. The S-matrix in the $ee$ subspace is
\begin{equation}\label{E:TripletSeeMatrix}
\mathbf{S}_{ee} =\sum_{k, p} t_k t_p |T_{k,p}\rangle \langle T_{k,p}|.
\end{equation}

\begin{figure}[thb]
\scalebox{1}{\includegraphics[width=\columnwidth]{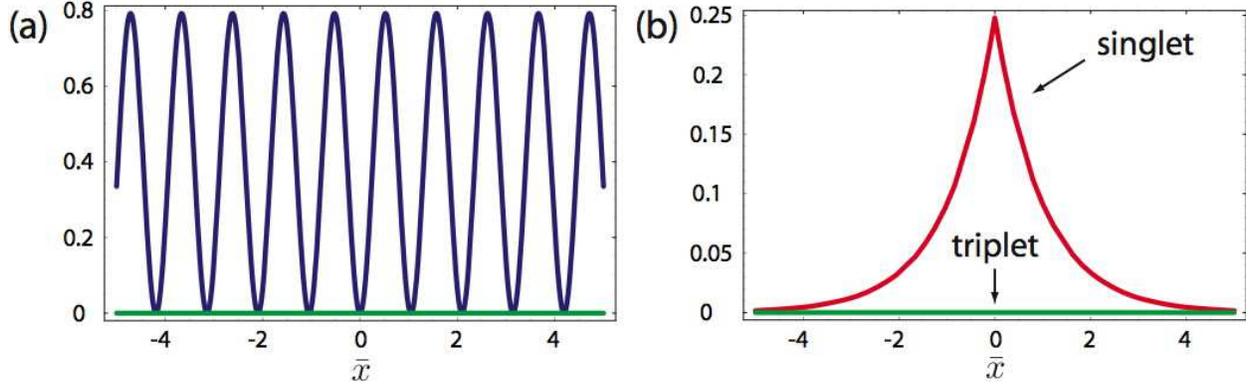}}
\caption{(color online). (a) Triplet $|t_2^{\text{\scriptsize t}}(x_1, x_2)|^2$ for $\delta E = 4\Gamma$ (blue curve) and $6\Gamma$ (green curve). $\Delta = -3\Gamma$. The triplet is always antibunched. (b) Singlet-triplet filtering. $\delta E = 6\Gamma = U = -2\Delta$. The singlet case ($|t_2^{\text{\scriptsize s}}|^2$) is denoted by the red curve and the triplet case ($|t_2^{\text{\scriptsize t}}|^2$) by the green curve. Both $|t_2^{\text{\scriptsize t,s}}|^2$ are normalized by $(\sqrt{2}/2\pi)^2$.}\label{Fi:triplet}
\end{figure}

For a triplet in-state with $S_z=0$,
\begin{equation}\label{E:InTripletRR}
|\Psi_\text{in}\rangle^{\text{\scriptsize t}} \equiv |T_{k_1,p_1}\rangle_{R\uparrow R\downarrow} =\int dx_1 dx_2 \,T_{k, p}(x_1, x_2) c_{R\uparrow}^{\dagger}(x_1) c_{R\downarrow}^{\dagger}(x_2) |\emptyset\rangle,
\end{equation}the corresponding wavefunction describing both electrons being transmitted is 
\begin{equation}\label{E:OutTripletRR}
t_2^{\text{\scriptsize t}}(x_1, x_2) =\bar{t}_k \bar{t}_p T_{k,p}(x_1, x_2) \propto \sin[\Delta(x_1-x_2)].
\end{equation} Thus the transport properties of the two electrons in triplet are the same as those of two independent electrons.  Fig.~\ref{Fi:triplet}(a) plots $|t_2^{\text{\scriptsize t}}(x_1, x_2)|^2$ for $\Delta = -3\Gamma$. When $\delta E = 6\Gamma = -2\Delta$, $t_2^{\text{\scriptsize t}}$ vanishes, since one electron is on resonance with the impurity and the corresponding single-electron transmission amplitudes in Eq.~(\ref{E:OutTripletRR}) vanishes.  

\emph{Singlet-triplet filter.} The different transport properties of the singlet and the triplet states allow one to construct a novel \emph{deterministic} filter that operates on the total spin angular momentum of the electron pair but not individual spins. In the triplet case, the transmission amplitude for the electron pair is zero when one electron is on resonance with the impurity with energy $\epsilon_d$. In the singlet case, the electron pair can go through the impurity by forming a bound pair of singlet state. Thus the filter deterministically generates a two-electron state that is entangled in spin, energy, and momentum space [Fig.~\ref{Fi:triplet}(b)]. The transmission amplitude attains maximum value when the singlet approximately satisfies both the single-electron and two-electron resonance conditions simultaneously.  

As a final remark, here we show that the electron pairing can be generated by strong on-site Coulomb repulsion. Since the same one-mode model also describes the S-wave scattering in two and three dimensions, we speculate that such a pairing mechanism could be important in bulk materials involving Anderson impurities. 

S. F. acknowledges financial support by the David and Lucile Packard Foundation.

\pagebreak
\newpage

\pagebreak
\newpage

\pagebreak
\newpage

\end{document}